\RequirePackage{fix-cm}
\documentclass{svjour3}                     
\smartqed  
\usepackage[table,xcdraw]{xcolor}
\usepackage{graphicx}
\usepackage{amsmath}
\usepackage[numbers]{natbib}
\usepackage{hyperref}
\usepackage{xurl}
\usepackage{caption}
\usepackage{rotating}
\usepackage{lscape}    
\usepackage{multirow, makecell, caption}

\journalname{Applied Intelligence}
\begin{document}

\title{FPSRS: A Fusion Approach for Paper Submission Recommendation System
}

\titlerunning{FPSRS: A Fusion Approach for Paper Submission Recommendation System}        

\author{Son T. Huynh \and
        Nhi Dang \and
        Dac H. Nguyen  \and
        Phong T. Huynh \and
        Binh T. Nguyen
}

\institute{Son T. Huynh \at
              AISIA Research Lab, Ho Chi Minh City, Vietnam\\
           \and
           Nhi Dang \at
              AISIA Research Lab, Ho Chi Minh City, Vietnam\\
            \and
           Dac H. Nguyen \at
              AISIA Research Lab, Ho Chi Minh City, Vietnam\\
            \and
           Phong T. Huynh \at
              AISIA Research Lab, Ho Chi Minh City, Vietnam\\
              \and
              Binh T. Nguyen (Corresponding Author) \at
              Vietnam National University in Ho Chi Minh City\\
              University of Science, Vietnam\\
              \email{ngtbinh@hcmus.edu.vn}}

\date{Received: date / Accepted: date}

\maketitle

\begin{abstract}

Recommender systems have been increasingly popular in entertainment and consumption and are evident in academics, especially for applications that suggest submitting scientific articles to scientists. However, because of the various acceptance rates, impact factors, and rankings in different publishers, searching for a proper venue or journal to submit a scientific work usually takes a lot of time and effort. In this paper, we aim to present two newer approaches extended from our paper \cite{huynh2021fusion} presented in the conference IAE/AIE 2021 by employing RNN structures besides using Conv1D. In addition, we also introduce a new method, namely DistilBertAims, using DistillBert for two cases of uppercase and lowercase words to vectorize features such as Title, Abstract, Keywords and then use Conv1d to perform feature extraction. Furthermore, we propose a new calculation method for similarity score for Aim \& Scope with other features; this helps keep the weights of similarity score calculation continuously updated and then continue to fit more data. The experimental results show that the second approach could obtain a better performance, which is $62.46\%$ and $12.44\%$ higher than the best of the previous study \cite{huynh2021fusion} in terms of the Top 1 accuracy.

\end{abstract}
\keywords{Recommender System, RNN, DistilBert}

\section{Introduction}
Nowadays, large-scale corporations are increasingly focusing on applying recommendation systems in human life. By diving into historical user data, these big companies can use recommendation algorithms to give the customers more excellent suggestions or understanding, enhancing users' experience and helping raise the profit of these companies. Hence, various organizations such as Google, Facebook, Amazon, eBay, Spotify, and Netflix have investigated human resources and money to improve and alleviate their recommendation algorithms in various company products.

Interestingly, various companies are constructing recommendation systems that can also be utilized in education. However, especially in the scientific academic, when junior scientists want to submit their scientific paper, they always wonder which publisher they should submit their research work. Consequently, these scientific papers are mistakenly submitted to journals or conferences, leading to rejection and wasting the time of both authors and reviewers. For this reason, we are motivated to study this problem to help the scientists, special master or postdoctoral, who have just stepped on the path of scientific research. They can easily submit their scientific work quickly and accurately with such applications. 

In the short version of this paper \cite{huynh2021fusion}, we proposed a recommendation system that, from the combinations of features such as Title, Abstract, and Keywords, could recommend for scientific researchers which journals of conferences they should submit to increase the chance of acceptance from publishers. We described a method using FastText as an embedding matrix and Convolution-1-dimension (Con1D) structure for extracting features from three inputs, including Title, Abstract, and Keywords. Additionally, we introduced a new feature: Aim \& Scope; we use FastText for a vectorized way and barely calculate similarity cosine by cosine without training. 

In this research, we enlarge our works by conducting an extensive investigation to replace Conv1D with other feature extractors like LSTM, BiLSTM, GRU, and BiGRU using Aims \& Scope and not using Aims \& Scope. Next, we propose a second approach using DistilBert as a pre-trained model instead of selecting previous feature extractors. We realize using DistilBert uncased without Aims \& Scope can gain accuracy in terms of Top1, Top3, Top5, Top10 are $0.5503$, $0.8398$, $0.8959$, and $0.9479$, respectively, in the dataset given in \cite{huynh2021fusion}. Whilst, using Aims \& Scope can enhance the accuracy terms of Top1, Top3, Top5, Top10 are $0.5537$, $0.8409$, $0.9010$, and $0.9524$, correspondingly. Furthermore, the model DistilBert cased can make the best results ever in two cases, including Aims \& Scope and without Aims \& Scope. The experimental results show that the proposed techniques can gain $0.5891$ in Top 1 accuracy in case of without using Aim \& Scope and $0.6246$ in Top 1 accuracy in case of using Aim \& Scope.

\section{Related Work}
Klamma and his colleagues \cite{original1} contributed one of the first datasets relevant to the paper submission suggestion problem in 2009. Extraordinary, this dataset is the combination of two data sources, DBLP\footnote{\url{https://dblp.org/}} and Eventseer\footnote{\url{https://www.eventseer.net/}}. Klamma et al. built a collaborative filtering model for their paper recommendation system by approaching crucial information about individuals who participated in related conferences.

Medvet et al. \cite{Medvet} studied various strategies for recommending relevant publication venues using abstract, title, and keywords as the input data and analyzing three methods (including Cavnar Trenkle, two-step Latent Dirichlet Allocation, and the combination between Latent Dirichlet Allocation and K-mean clustering). They investigated in the dataset that belonging to 300 conferences, including 5800 papers in computer science from Microsoft Academic Search\footnote{\url{https://academic.microsoft.com/}}.
After that, Luong et al. \cite{luong1} proposed a method to extract suitable features that use social network analysis and co-author networks. Finally, this study compared the performance on one dataset having 960 documents belonging to 16 ACM conferences.

Due to its essential role, the publication recommendation system is a crucial topic for both researchers and publishers. IEEE, Springer, or Elsevier are familiar publishers who provide those practical systems for researchers finding suitable venues for their works. With a common approach, those systems take these typical features, including title, abstract, and/or keywords (or main topics) of submission as input and list out all top matched conferences or journals. However, those publishers' recommendation systems only provide a list of conferences and journals that already exist on their database. This action makes one step back on the progress of publishing or sharing researchers' ideas. Besides production level systems, many research works related to the submission recommendation have been published in recent years. For instance, in 2018, Wang and colleagues \cite{WANG20181} showed promising performance with an accuracy of 61.37\% in recommending top proper conferences and/or journals with a given manuscript. Wang used Chi-square and TF-IDF as feature engineering layers and a linear regression model for the classifier. Later, with the same data, with a simple deep learning approach \cite{ieaaie2020}, Son et al. outweighed the performance of Wang's approach by using MLP (Multi-layer perceptron) as a classifier instead of using logistic regression with accuracy (Top 3) of 89.07\%. 
In 2020, Dac et al. \cite{Sofsem} used a new technique for this problem by investigating numerous deep learning methods as LSTM \cite{lstm}, GRU \cite{gru}, Conv1D, and the ensemble method. Interestingly, the experimental results \cite{Sofsem} could outperform the previous results \cite{ieaaie2020} in terms of Top 1, Top 3, Top 5, and Top 10 accuracies. However, the dataset's volume that Wang used has only 14012 samples, which is not large enough to be reliable or high confident results. 

In 2021, Son et al. released a new dataset for this problem \cite{huynh2021fusion}; this dataset has 414512 scientific papers from Springer's publisher. Significantly, they used Aims and Scopes as an extra feature for increasing the performance of the paper submission system. As a result, they could gain the best results in Top 1, Top 3, and Top 5 accuracies, which are $0.5002$, $0.7889$, $0.8627$, respectively.

\section{Background}
This section introduces several basic concepts related to the main problem used in our proposed techniques and experiments later. 

\subsection{FastText}
Piotr Bojanowski and his colleagues first introduced FastText in 2016 \cite{fasttext}. According to the authors, FastText is an extension of word2Vec, and this vectorization can fix a significant drawback of word2vec because it can only use words in the dataset. FastText use a second approach by dividing the text into small chunks called n-grams for each term. For instance, ``avocado'' would be ``avo'', ``voc'', ``oca'', ``cad'', and ``ado'' the vector of the word ``avocado'' would be the sum of all these, instead of training for word units in wor2vec. Therefore, it handles very well for rare word cases.

\subsection{DiltilBert}
DistilBert \cite{ditlbert} is a variation of Bert \cite{bert}. According to the authors of this paper, DistilBert is lighter and faster at inference time than Bert while able to reach similar performances on many downstream tasks with knowledge distillation. Because of requiring a smaller computational training budget, DistilBert can train and apply on compact devices with less strong hardware power.

DistilBert use a mechanism is called distillation \cite{distilation1}\cite{distilation2}. This technique constructs one model that plays a role as the student is used to reproduce the behavior of a massive model or an ensemble of a model called the teacher. According to the authors, DistilBert can retain 97\% predictive efficiency but only uses half the downstream task parameter.
This paper uses DistlBert as a pre-train model and fine-tune it with our data in two cases, including lower and upper words.
\subsection{RNN}
Recurrent Neural Network (RNN) \cite{rnn1} \cite{rnn2} is a neural network that processes information in sequence/time-series, with preprocessing of ordered data.
\subsubsection{LSTM $\&$ BiLSTM}
LSTM \cite{lstm} is a special architecture of RNN \cite{rnn1}\cite{rnn2} capable of learning long-term dependencies introduced by Hochreiter \& Schmidhuber in 1997. LSTM consist of three gates:
\begin{itemize}
 \item[*] Forget gate: This gate is responsible for deciding how much word to receive before the cell state.
 \item[*] Input gate: This gate play a role decide how much to take from the input of the state and the hidden layer of the previous layer.
 \item[*] Update gate: This gate decide how much to take from the cell state to be the output of the hidden state.
\end{itemize}
 
LSTM still has a vanishing gradient phenomenon but less than RNN. Moreover, when carrying information on cell state, it is seldom necessary to forget the previous cell value. Thus, this architecture has been popular and widely used because of its advantages compared to RNN. 

BiLSTM is the variation of LSTM; this model is constructed by stacking two LSTM models: one receives data in a forward manner, while the other receives data in a backward way. BiLSTMs dramatically raise the amount of information possible to the network, enhancing the context available to the model.

\subsubsection{GRU $\&$ BiGRU}
Introduced by Cho et al. \cite{gru} in 2014, GRU aims to solve the gradient vanishing problem that accompanies RNN. GRU is considered a variant of LSTM as both are designed similarly. In some cases, the results may be equally good. GRU includes two gates: 
\begin{itemize}
 \item[*] Update gate: This gate decide how much past information to forget.
 \item[*] Reset gate: This gate decides what information to throw away and what new information to add.
\end{itemize}
Because of the less complex structure when compared to LSTM, the training time of GRU is faster, and the performance is not too bad, so the GRU is now widely used.
Similar to BiLSTM, BiGRU is also built by stacking two GRU layers.
\subsection{Convolution1D}
Since the first impressive official appearance in 1998, besides the Gradient Descent for Multilayer Neural Networks that mentioned in Yann Lecun et al. \cite{726791}, Convolutional Neural Network emerged and developed as one of the most popular deep learning models in Computer Vision. Convolutional layer, nowadays, is an integral layers of many typical foundation state-of-the-art models' architectures like VGG \cite{vgg}, Resnet \cite{resnet}, Inception Net (also known as GoogLeNet) \cite{DBLP:journals/corr/SzegedyIV16} and SqueezeNet \cite{iandola2016squeezenet}.

Convolutional neural networks are space invariant artificial neural networks (SIANN). Intuitively, the convolutional layer uses kernels or filters to slide along the input features and project them on another "meaningful" dimension created hidden feature maps. The networks can precisely extract or understand the original input's semantic meaning by taking the "hidden feature maps" as input for the following layers. Because the kernels (filters) of a convolutional layer scan multiple cells at once, it can express the hidden meaning of the cells and their effects on each other.
Many researchers try to apply the Convolutional layer according to its adjacent cell extraction ability and its fast inference time compared to the RNN layer (event LSTM or GRU). It is known as Conv1D (a one-dimensional version) into their models' architectures for Natural Language Processing tasks such as \cite{kim2014convolutional}, \cite{zhang2016characterlevel}.

\subsection{Cosine Similarity}
There are several to calculate the similarity between two documents. In this paper, we choose a primary method as a baseline, which is the cosine similarity. The basis is converting content to vector format by embedding method. We can get pre-trained embedding weight from several websites like Word2vec, Fasttext, etc.

In each document, we calculate the center of the text, which means we calculate the mean vector of all embedding vectors representing a word. With two center vectors $A$ and $B$ of two papers, we calculate similarity by the formula below:
$$similarity=\dfrac{A.B}{||A||.||B||}=\dfrac{\displaystyle\sum^n_{i=1}A_iB_i}{\sqrt{\displaystyle\sum^n_{i=1}A_i^2}.\sqrt{\displaystyle\sum^n_{i=1}B_i^2}},$$
where $n$ is the dimension of these two vectors.

\section{Our Methodology}
This section presents a second approach for a paper recommendation system, from a scientific article with a Title, Abstract, and list of Keywords. In addition, we propose a further use of Aims \& Scope as a new feature for enhancing the performance of the paper recommendation system. Our system allows helping the scientists able to gain more helpful information about which venues suit their work by recommending top $N$ journals or conferences. This system can have potential applications and become a vital tool for the scientific community. In what follows, we show how to enhance the old approach by using state-of-the-art techniques. We will illustrate three approaches, including the baseline method,  the first approach, and our second approach. In addition, for each approach, we describe in detail how to preprocess the data.

\subsection{The baseline method}
It is the baseline technique that was applied in previous our paper \cite{huynh2021fusion} presented at IEA/AIE 2021.
\subsubsection{Data Preprocessing}
For constructing the baseline approach, we show the data processing step by step on our data to computers able to comprehend and efficiently utilize embedding methods to extract valuable features:
\begin{enumerate}
 \item{We lowercase all text to convert all words returning to the same format.}
 \item{We eliminate not-be-alphabet words containing trivial semantic to the problem (for instance, a word "pre-treatment" or an email "author@gmail.com").}
 \item{We remove single letters that likely do not have pretrained weights in well-known pretrained word vectors like FastText Common Crawl\footnote{https://fasttext.cc/docs/en/english-vectors.html}.}
  \item{We remove words within stopwords downloaded from the \textit{Natural Language Toolkit} (NLTK\footnote{https://gist.github.com/sebleier/554280}) and additional stopwords we define.}
  \item{We remove unnecessary space from the beginning to the end of the text after doing four steps above.}
\end{enumerate} 

\subsubsection{Modeling}
We use FastText as an embedding for seven types of input including Title (T), Abstract (A), Keywords (K), Title + Abstract (T+A), Title + Keywords (T+K), Abstract + Keywords (A+K), Title + Abstract + Keywords (T+A+K). After that, we use Convolution 1D as a feature extractor (This feature we name $extracted\_feature$) and connect with using three blocks that each block consists of one Fully Connected layer and Dropout layer. Next, we utilize Softmax as a classifier for predicted relevant journals or conferences from a scientific paper. Interestingly, we use Aims \& Scope as an extra feature to increase the performance by using FastText for embedding the Aims \& Scope. We then use Cosine Score to calculate the similarity between Aims \& Scope that belong to journal or conference and the features. This similarity score will be concatenated with $extracted\_feature$. Finally, we pass this concatenated feature through three blocks, FullyConnected and Softmax, for constructing an appropriate classifier. One can see more detail in our paper \cite{huynh2021fusion}.

\subsection{Our First Approach}

\begin{figure}[t]
    \centering
    \includegraphics[scale=0.3]{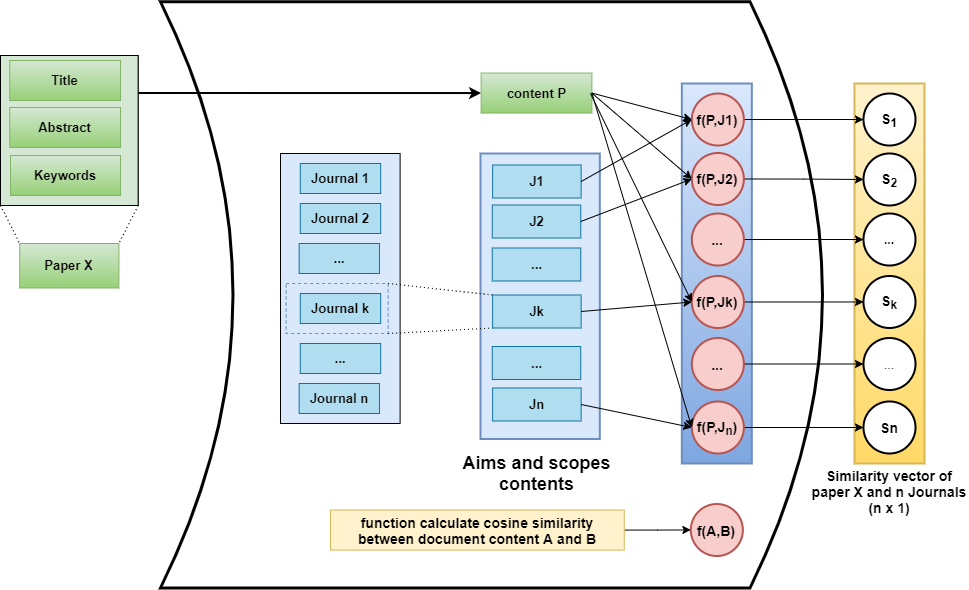}
    \caption{Similarity extraction model with aims\&scope}
    \label{fig: Similarity calculate model}
\end{figure}

\begin{figure}[t]
    \centering
    \includegraphics[scale=0.25]{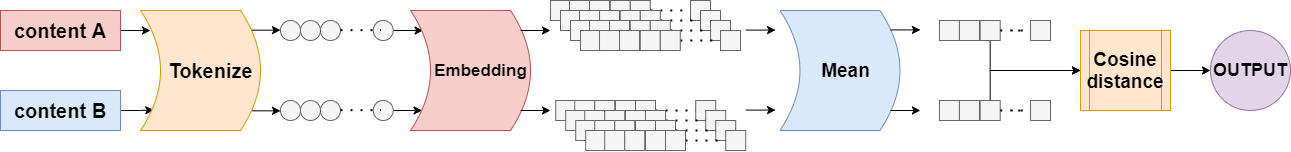}
    \caption{Cosine distance calculation}
    \label{fig: Cosine distance calculation}
\end{figure}

\begin{figure}[ht!]
    \centering
    \includegraphics[scale=0.32]{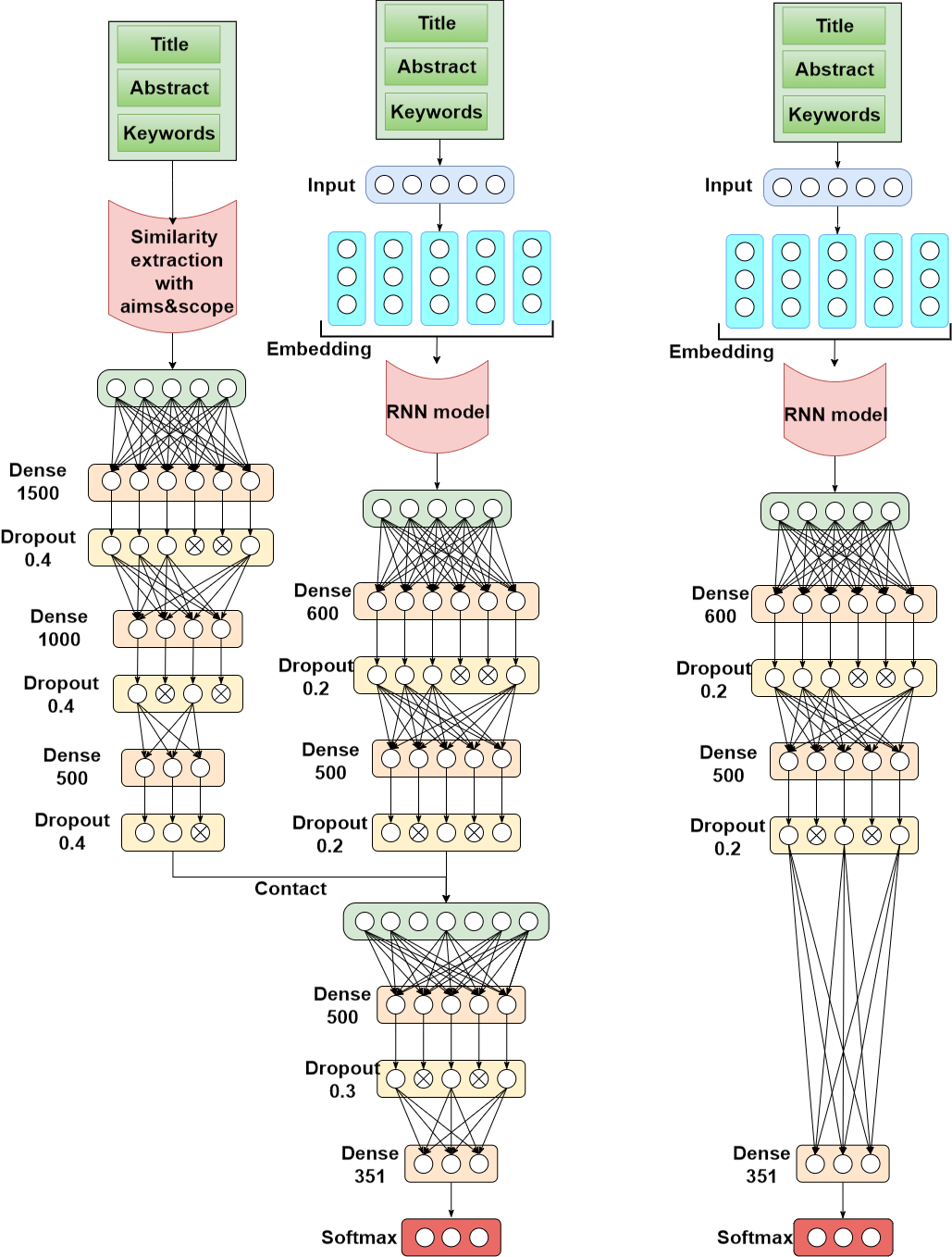}

    \caption{Model with similarity extraction (left) Model without similarity extraction with aims\&scope (right)}
    \label{fig: old approach}
\end{figure}

Improving the performance that Dac and his colleagues proposed in \cite{Sofsem}, we \cite{huynh2021fusion} contributed to this method by adding Aims and Scope to increase the performance compared to the preceding works. However, we realize that both approaches barely used Convolution1D as a feature extractor. As a result, in this paper, we aim to replace Convolution1D with further advanced models for sequential data such as LSTM \cite{lstm}, BiLSTM, GRU \cite{gru}, and BiGRU.

\subsubsection{Data Preprocessing}
As a result of inheritance of past studies, we reuse the preprocessing techniques that Son et al. \cite{huynh2021fusion} used in their paper. However, we add further steps to enhance the performance of the proposed methods. We describe the preprocessing steps as follows:
\begin{enumerate}
\item First, we remove numeric characters and all characters which are not in the alphabet (even in case the number written in words like fifteen, six, seven, sixteen, etc. could be listed in our defined stopwords)
\item Second, we separate contacted words in case the lowercase letters stand before the uppercase letter (for instance, ArtificialIntelligence $\rightarrow$ Artificial Intelligence)
\item Next, we remove all crawling errors found in the relevant texts.
\item Then, we lowercase every word to decrease complexity.
\item Finally, we remove all English stopwords (downloaded using NLTK library in Python), our defined stopwords, and all redundant backspaces.
\end{enumerate}

\begin{table}[]
    \centering
    \renewcommand{\arraystretch}{1.3}
    \caption{Preprocessing example}
    \scalebox{0.9}{
    \begin{tabular}{|c|c|}
    \hline
        Before & After \\ 
    \hline
    \multirow{3}{*}{\begin{tabular}[c]{@{}l@{}}Fifteen lambs (av. BW, 22.5$\hat{a} \backslash x80 \backslash x89 $ \\
    $\hat{A} \pm \hat{a} \backslash x80\backslash x891.6 \hat{A}\backslash xa0kg$) were \\  randomly allotted into 3 treatments \\ \end{tabular}} & \multirow{2}{*}{\begin{tabular}[c]{@{}l@{}} lambs av bw kg randomly \\  allotted treatments \end{tabular}} \\
     & \\
     & \\\hline
    Einstein$\hat{a}\backslash x80\backslash x93$Podolskiy$\hat{a}\backslash x80\backslash x93$Rosen & einstein podolskiy rosen \\\hline
     simultaneous-approximation-term & simultaneous approximation term \\\hline
     \multirow{6}{*}{\begin{tabular}[c]{@{}l@{}} In this work, a theoretical approach was  \\ developed for modelling olefins diffusion in two \\ typical zeolites, HZSM-5 and HSAPO-34.  \\ Activation barrier between large cavities and \\channels was determined using \\ Lennard$\hat{a}\backslash x80\backslash x93$Jones (LJ) potentials\end{tabular}}
     &  
     \multirow{6}{*}{\begin{tabular}[c]{@{}l@{}} work theoretical approach developed  \\   modelling olefins diffusion typical  \\ zeolites hzsm hsapo activation barrier \\ large cavities channels determined using \\ lennard jones lj potentials \end{tabular}}
    \\
     & \\
     & \\
     & \\
     & \\
     & \\\hline
    
    \end{tabular}
}
    \label{tab:my_label}
\end{table}

\subsubsection{Modeling}
Our previous approach separated our model architecture into three flows: the main flow, the similarity flow, and the concatenation flow. The final fully-connected layer for output contains the number of units in the hidden layer as the number of considered journals and conferences: The main flow, the similarity flow, and the concatenation flow.

Firstly, we want to measure the model's performance without using Aim \& Scope as an extra feature. Instead, this model takes input data combined with Title, Abstract, and Keywords. After that, we utilize FastText to vectorize a text into an embedding matrix having the dimension as $M \times 300$. Here, $M$ is the number of words in a text, and if the actual word count of a text is not enough $M$, we pad the sentence after so that the length is equal to $M$. We then feed this embedding matrix through each of four models, including LSTM, BiLSTM, GRU, and BiGRU, with 100 units to efficiently extract information from low to high levels in the experiments.
Furthermore, we use two blocks where each block includes a fully-connected layer with ReLu as activation and a Dropout layer with a dropout rate of 0.2. We name this flow as $main\_flow$ and use Softmax at the end of the model for the final classification. One can see more detail of this model on the right-hand side of Figure \ref{fig: old approach}.

Secondly, we add a module that can extract the matching score of a given paper with every considered journal or conference. As described in the left-hand side of Figure \ref{fig: Similarity calculate model}, each input data contains the title, the abstract, and the list of keywords from the chosen paper. One can use them to calculate the matching score between the selected article and every journal or conference stored in the database. After computing the feature vector with the length of the number of considered journals and conferences, one can pass that vector sequentially through a fully-connected layer of 1500 units, a fully-connected layer of 1000 units, and the last fully-connected layer of 500 units. We also choose a RELU activation function and a dropout layer of the rate 0.4 behind each fully-connected layer in this flow. We then concatenate this flow with $main\_flow$ as shown in the left-hand side of Figure \ref{fig: Similarity calculate model}.  After that, we pass this concatenated flow through a fully-connected layer of 500 units and a dropout layer of the rate 0.3. Finally, we use Softmax to rank the top journals or conferences relevant to a given scientific paper.

As depicted in Figure \ref{fig: old approach}, we do experiments for this approach by measuring two different models for the main problem. In the first model, we combine three flows using aims\&scope extraction model as shown in Figure \ref{fig: Similarity calculate model}. We only use the main flow for the second model without applying the aims\&scope extraction model for better comparison.

\subsection{Our Second Approach}
This section illustrates our second approach for the paper submission recommendation system in detail. 

We utilize DitilBert \cite{ditlbert} as a embedding layer for vectorizing input instead of using FastText in previous works \cite{Sofsem}, \cite{huynh2021fusion}. Moreover, we use Convolution1D with different kernel sizes with the hope that the model can extract various features from which to make the system superior from various kernel sizes and initialization weights. Interestingly, to take advantage of feature Aim \& Scope, we also apply DistilBert for Aim \& Scope and calculate the similarity score between  Aim \& Scope and other features. We name this model that applies DistilBert for Title, Abstract, Keywords, and Aim \& Scope as DistilBertAims.

\subsubsection{Data Preprocessing}
Semantic understanding words of DistilBert, we do not reuse the data processing step as mentioned above in the first approach. In our study, the manuscript's abstract regularly includes mathematical equations or Latex scripts, which cause various problems in feature selection. Consequently, we eliminate all phrases likely to be Latex codes or mathematical equations from a given abstract.

After cleaning data, we split each word in the input into subwords to alleviate the burden of extended English vocabulary. Then, we add two tokens [CLS] in the head and [SEP] in the end in each list of words. Finally, for each different type of feature combination, there is a different number of token sequence lengths, as shown in Figure \ref{tab: maxlength}.
\begin{table}
\centering
\renewcommand{\arraystretch}{1.2}

\caption{Max Sequence Length of each combination of feature}
\begin{tabular}{lc}
\hline
Features                & Max Sequence Length  \\ 
\hline
Title                   & 128                  \\ 
\hline
Abstract                & 512                  \\ 
\hline
Keywords                & 128                  \\ 
\hline
Abstract+Keywords       & 512                  \\ 
\hline
Title+Keywods           & 256                  \\ 
\hline
Title+Abstract          & 512                  \\ 
\hline
Title+Abstract+Keywords & 512                  \\
\hline
            
\end{tabular}
\label{tab: maxlength}
\end{table}

\subsubsection{Modeling}
In this method, we design a model to get two inputs; one is a combination of three features, including Title, Abstract, and Keywords ($input_1$), the other is Aims \& Scope ($input_2$). Therefore, in this model, we have two phases: One is the extraction step for $input_1$, the other is the step that calculates the similarity of $input_1$ and $input_2$. Both phases are described in the following details:

First of all, the DistilBert model gets $input_1$ including Title, Abstract, and Keywords and outputs all hidden state of the model as an embedding matrix that presents the information of $input_1$. In addition, we set the dimension of the word vector as $768$. Next, this embedding matrix passes through 3 Convolution1D layers in parallel with kernel sizes are $2\times 2$, $3 \times 3$, and $4 \time 4$ respectively, we choose the number of filters in each Convolution1D layer is $200$. After that, we reduce the dimension of each output after using the convolution operator by utilizing the GlobalMaxPooling layer; we name them $maxpool_1$, $maxpool_2$, $maxpool_3$, respectively. 

We denote $input_2$ as Aims \& Scope; this input includes $351$ documents, equal to the number of journals in our database. As we described above, we calculate the cosine score of two inputs which are $input_1$ and $input_2$, as follows: firstly, two inputs also share weights in the model DistilBert, and likewise, this model is used in phase one. However, what is different here is the DistilBert model barely outputs the last state instead of all states as phase one. As a result, the output of $input_1$ after passing through DistilBert is a vector having the dimension of $768$. While the output data of the remaining input is the matrix having a size of $351 \time 768$, the number $351$ is the number of venues that we mentioned above. After that, we compute the cosine score of this vector and matrix, and the result is the vector has the size is $1 \times 351$, we denote this vector as $cosine\_score$. This method gets ideas from Siamese Neural Network \cite{siamese1}\cite{siamese2}, by backpropagation during training, the weights of the last input of DistilBert are updated then our system will learn which input pairs are close together.

\begin{figure}[ht!]
    \centering
    \includegraphics[scale=0.5]{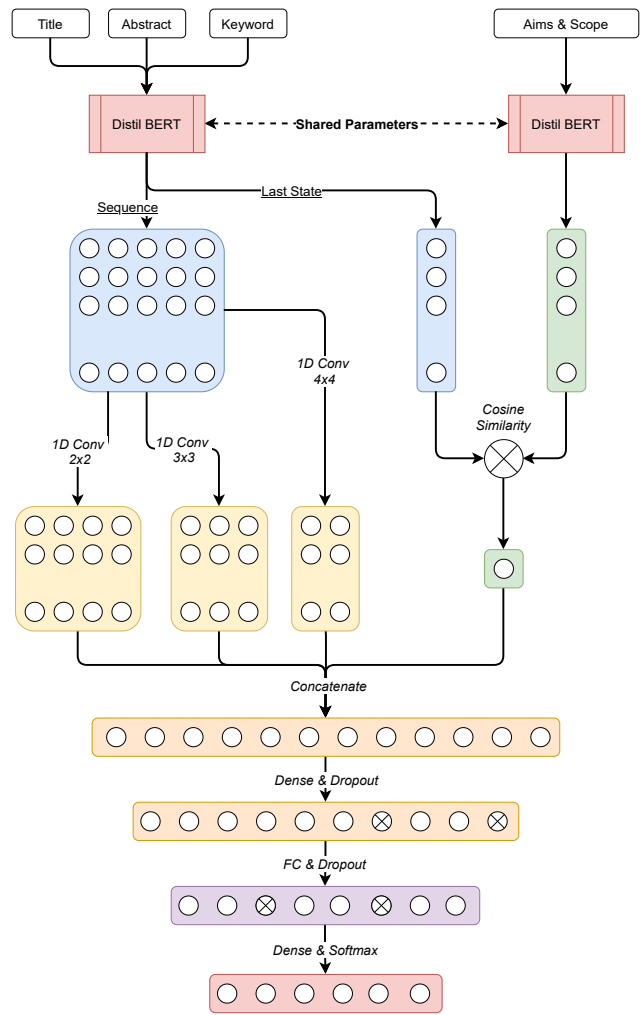}
    \caption{The architecture of our second approach using DistilBert}
    \label{fig: bert}
\end{figure}

Finally, we concatenate three vectors that contain the feature of data to make only one unique feature vector and pass it through two blocks. Each block accommodates one fully-connected layer with the number of nodes is $n$, and one Dropout layer with a rate of $0.2$, and the activation function is Relu. These two blocks have the number of nodes in the fully connected layer as $500$ and $400$, respectively. Ultimately, we use Softmax to compute the matching score between each journal and the selected scientific paper . One can see comprehensively in Figure \ref{fig: bert}.

In addition, the model that does not use Aim \& Scope is built by removing the similarity calculation between feature combinations and Aim \& Scope.

\section{Experiments}
We measure the performance of baseline models and our models on a computer with Intel(R) Core(TM) i9 9900K with eight cores, 16 threads running at 3.6GHz with 64 GB of RAM, and an Nvidia GeForce RTX2080Ti GPU.

We use Numpy and Pandas as packages for processing and reading data in our experiments. We employ the Regex package as the primary tool to clean and normalize data in the data processing phase. Furthermore, as our dataset is enormous, we need to use a Multiprocessing package to utilize the CPU's multithreading capabilities. Thus, it makes our model run more quickly. Finally, we use Keras as a main API for the modeling procedure.

\subsection{Datasets}
We comprehensively describe the dataset used in the baseline approach and two new methods for our experimenting in what follows.

During this research, for comparing the performance of our proposed approach with previous work, we utterly experiment on the dataset that is submitted first time by Son et al. \cite{huynh2021fusion}. This dataset consists of scientific papers collected from the publisher Springer\footnote{\url{https://www.springer.com/gp}}.
According to Son and colleagues, after the data collection process, this dataset has 414512 papers. To facilitate future comparison, the authors divided the dataset into three sets, including training set, test set, and validation set, where the splitting ratio is $60\%/20\%/20\%$. In other words, this dataset consists of 248707 samples for the training dataset, 82902 articles for the testing dataset, and 82,903 for the validation dataset.

Each scientific paper has three features, including Title, Abstract, Keywords. One can see holistically in Figure \ref{tab: sample}. Especially, Son et al. \cite{huynh2021fusion} crawled the corresponding aims and scopes in Springer website for all journals. We measure the performance among selected models on existing features like Title, Abstract, Keywords and combine the aims and scope. One can observe the sample of aim and scopes as shown in Figure \ref{tab: sample_aims}.
\begin{table}[ht]
\centering
\caption{Several samples collected with the corresponding title, abstract, and keywords.}
\renewcommand{\arraystretch}{1.3}
\label{tab: sample}
\begin{tabular}{|l|l|l|}
\hline
Title                                                                                                                                                                                          & Abstract                                                                                                                                                                                                                             & Keywords                                                                                                                      \\ \hline
\begin{tabular}[c]{@{}l@{}}Prediction of mechanical \\ and penetrability properties \\ of cement-stabilized clay \\ exposed to sulfate attack by \\ use of soft computing methods\end{tabular} & \begin{tabular}[c]{@{}l@{}}The authors describe \\ a novel sensor for \\ chlorogenic acid (CGA) \\ detection/quantification \\ in food samples. \\ The photosensor is \\ based on a composite of \\ titanium dioxide...\end{tabular} & \begin{tabular}[c]{@{}l@{}}Photosensor, TiO2,  \\ Acridine orange, \\ Chlorogenic acid\end{tabular}                           \\ \hline
\begin{tabular}[c]{@{}l@{}}Amperometric Photosensor \\ Based on Acridine Orange/TiO\end{tabular}                                                                                               & \begin{tabular}[c]{@{}l@{}}Similar to its effects on \\ any type of cementitious \\ composite, it is a well-\\ known fact that sulfate \\ attack has also a negative \\ influence...\end{tabular}                                    & \begin{tabular}[c]{@{}l@{}}Cement-stabilized soil, \\ Strength, Penetrability, \\ BPNN, ANFIS, \\ Soft computing\end{tabular} \\ \hline

\end{tabular}
\end{table}

\begin{table}[]
\centering
\renewcommand{\arraystretch}{1.3}
\caption{Several examples collected with the corresponding aims and scopes in the relevant journals.}
\label{tab: sample_aims}
\begin{tabular}{|l|}
\hline
\multicolumn{1}{|c|}{Aims and Scope}                                                                                                                                                                                                                                                                                           \\ \hline
\begin{tabular}[c]{@{}l@{}}Iranian Journal of Science and Technology, Transaction A, Science (ISTT) is devoted to \\ significant original research articles of moderate length (not more than 20 pages in the \\ ISTT format), in a broad spectrum of Biology, Chemistry, Geology, Mathematics, and \\ Physics...\end{tabular} \\ \hline
\begin{tabular}[c]{@{}l@{}}Microfluidics and Nanofluidics is an international peer-reviewed journal that aims to \\ publish papers in all aspects of microfluidics, nanofluidics and lab-on-a-chip science \\ and technology....\end{tabular}                                                                                  \\ \hline
\end{tabular}
\end{table}

\subsection{Evaluation Metrics}
In our recommendation system, we use $Accuracy@K$ as a primary metric for performance our proposed approaches. We define the $Accuracy@K$ according to the following mathematical formulas:
\begin{equation*}
    Accuracy@K_i=\frac{TP@K_i+TN@K_i}{TP@K_i+TN@K_i+FP@K_i+FN@K_i}
\end{equation*}
\begin{itemize}
    \item $K$ is the top of $K$ recommended results in class $i^{th}$.
    \item $TP@K$ is the number of samples actual True which is predicted True allows top of $K$ recommended results in class $i^{th}$ .
    \item $TN@K$ is the number of samples actual True which is predicted False allows top of $K$ recommended results in class $i^{th}$.
    \item $FP@K$ is the number of samples actual False which is predicted True allows top of $K$ recommended results in class $i^{th}$.
    \item $FN@K$ is the number of samples actual False which is predicted False allows top of $K$ recommended results in class $i^{th}$.
\end{itemize}
\begin{equation*}
    Accuracy@K=\frac{\sum_{i=1}^NAccuracy@K_i}{N},
\end{equation*}
where N is the number of Journals in the dataset.

\subsection{Experimental results}
In this section, we compare and analyze the performance of our proposed technique (DistilBertAims) which has architecture as Figure \ref{fig: bert} (Table \ref{tab: result2}) with the RNN-based models' (like LSTM, BiLSTM, GRU, and BiGRU) and Son's model - \cite{huynh2021fusion} which has architecture as Figure \ref{fig: old approach}.

Firstly, the first approach using variations of RNN including LSTM, BiLSTM, GRU, BiGRU shows that the result is 
almost equal compared to the baseline method in terms of top-1, top-3, top-5, and top-10 accuracy. Next, using Aim \& Scope as an additional feature also shows that the result of using this feature is always higher than not using it. This result can prove that using Aims \& Scope as an extra feature can help to yield good results in most cases.

At a glance,according to Tables \ref{table: old_result}, \ref{tab: result1}, \ref{tab: result1_part2}, \ref{tab: result2}, the performances in top-k accuracy of running models on the entire feature combination (TAKS) are very stable and the highest in most cases compared to other combinations. Both DistilBertAims and Recurrent-based neural networks' top-1 accuracy when processing on TAKS combination (uncased) are higher than different feature combinations (above 48\%). Especially, with DistilBertAims as a model, taking TAKS (cased) as input's performance dominate any other feature combinations as well as any models in most top-k accuracy (k=1,3,5,10) with 62.46\%, 90.32\%, 94.89\% and 97.96\% respectively - Table \ref{tab: result2}. 

Having Aim \& Scope in the feature combination, models can extract some implied "nugget" patterns. Note that models show their better, more precise predictions in almost feature combinations having S (Aim \& Scope) - top-k accuracy of having S combinations are higher than others without S, about 3\% overall. And with DistilBertAims as the backbone, Conv1D (multi-size kernels) as filters, adjacent with Siamese convention for the similarity score, DistilBertAims precisely categorizes the proper conferences/journals for the corresponding contextual input and generally achieves better predictions compared to the Son et al. with the highest excess efficiency for top-1 accuracy of over 7\% (the uncased one). 

Finally, we expanded our experiments to the contextual feature differences by applying DistilBertAims on two different inputs, the cased approach (keep the upper-cases) and the uncased approach (lower-cased). This exploration gives us a fantastic outcome; the overall performance of DistilBertAims on cased feature combinations outperforms all other models on all different feature combinations, typically better than our DistilBertAims on uncased from 3 to 10\% in all top-k accuracy.

In short, the DistilBertAims model can extract the implied information of the contextual input much more than typical RNN based models and standalone transformers by combining advantages of Transformer architecture, data filtering capabilities of CNN layer, and the similarity extracted from Siamese structure. Moreover, with Aim \& Scope as an additional input, DistilBertAims shows its dominant and stable performance in solving this problem.

\begin{figure}[ht!]
    \centering
    \includegraphics[scale=0.3]{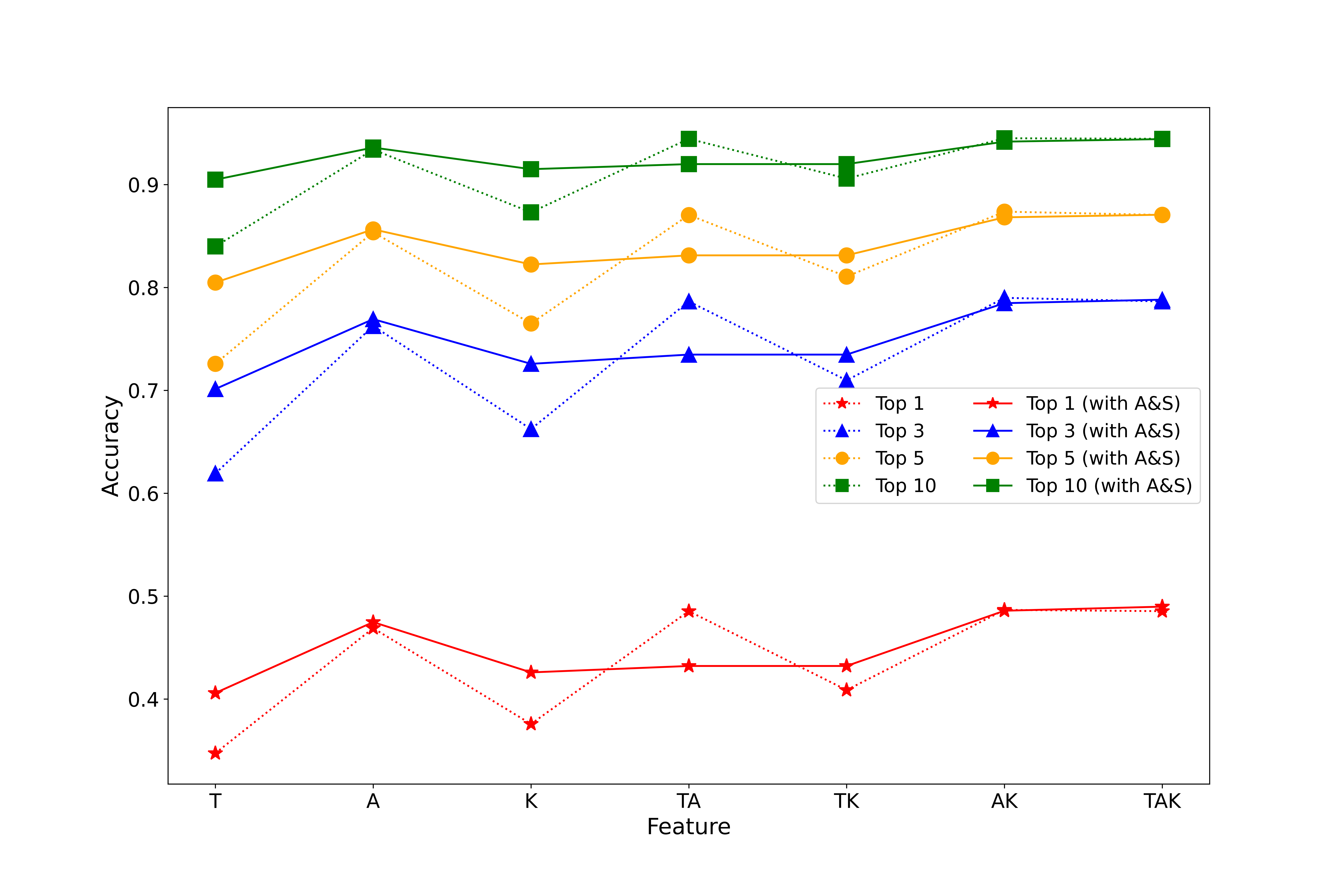}
    \caption{The performance of different features for the first approach using GRU as a feature extractor. Here, we compare the performance by Accuracy@K (K = 1, 3, 5, 10) for two cases: without/with using “Aims and Scopes”}
    \label{fig: gru}
\end{figure}

\begin{figure}[ht!]
    \centering
    \includegraphics[scale=0.3]{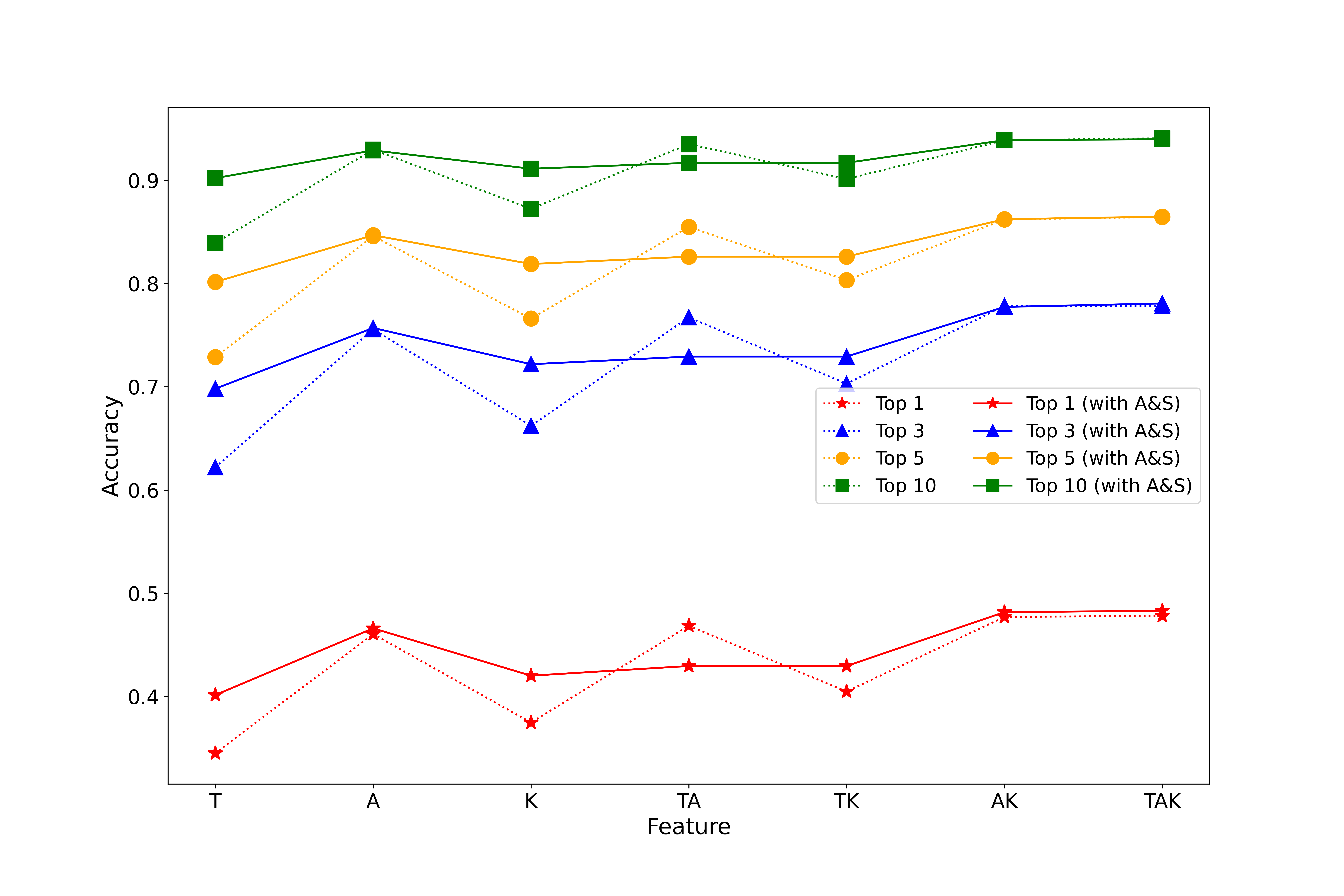}
    \caption{The performance of different features for the first approach using BiLSTM as a feature extractor. Here, we compare the performance by Accuracy@K (K = 1, 3, 5, 10) for two cases: without/with using “Aims and Scopes”}
    \label{fig: bilstm}
\end{figure}

\begin{figure}[ht!]
    \centering
    \includegraphics[scale=0.3]{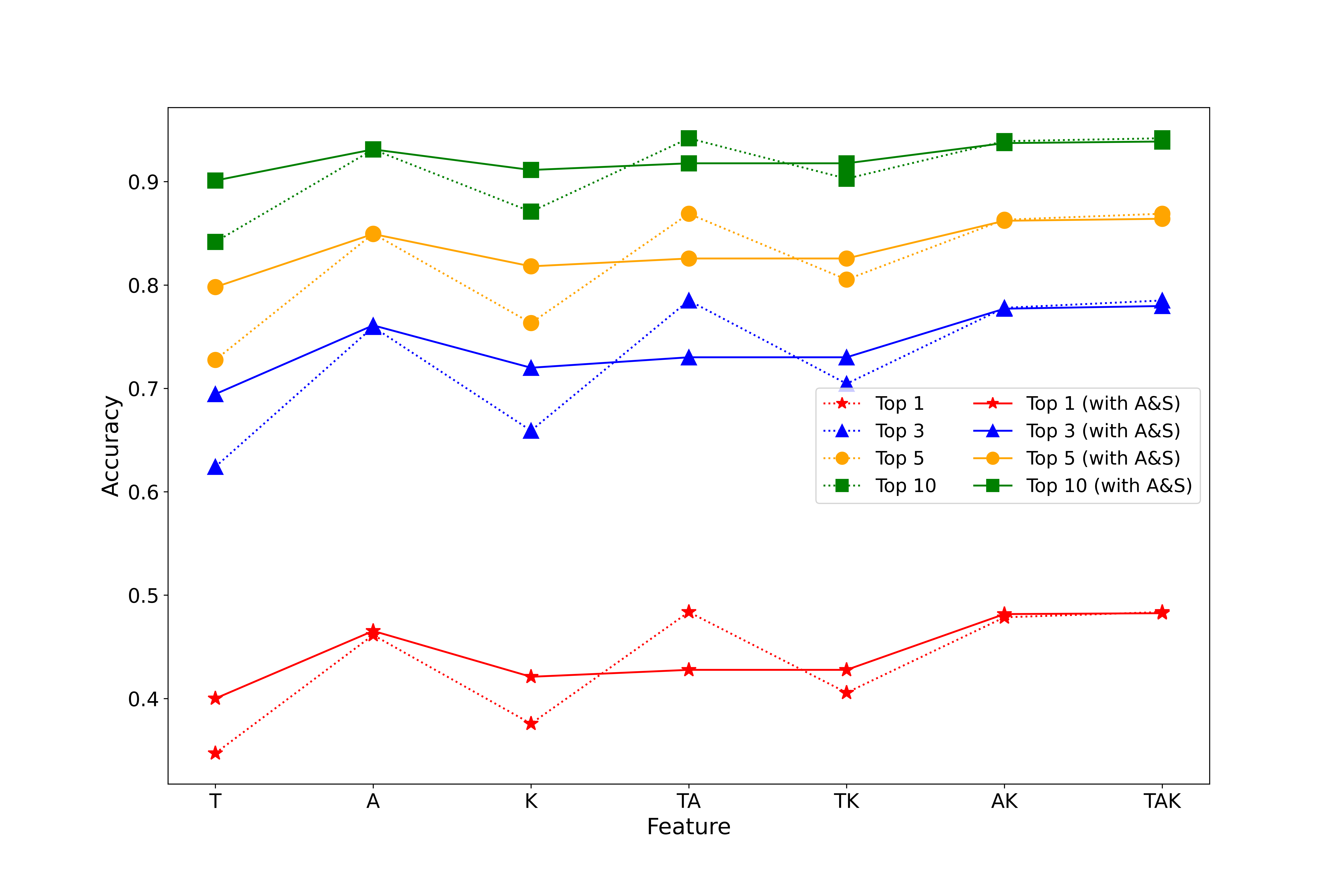}
    \caption{The performance of different features for the first approach using LSTM as a feature extractor. Here, we compare the performance by Accuracy@K (K = 1, 3, 5, 10) for two cases: without/with using “Aims and Scopes”}
    \label{fig: lstm}
\end{figure}

\begin{figure}[ht!]
    \centering
    \includegraphics[scale=0.3]{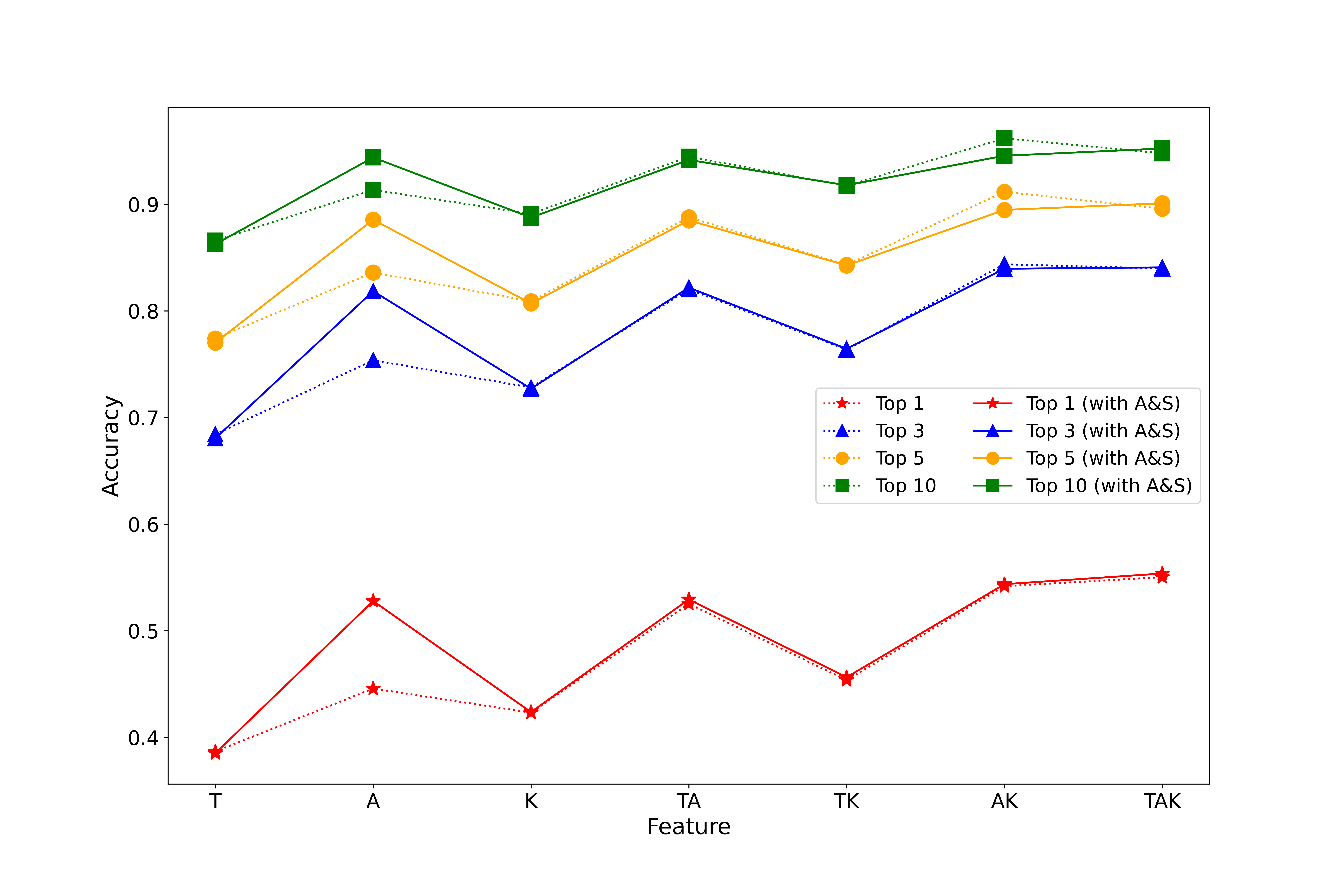}
    \caption{The performance of different features for the second approach that uses DistilBert uncased. Here, we compare the performance by Accuracy@K (K = 1, 3, 5, 10) for two cases: without/with using “Aims and Scopes”}
    \label{fig: bert_uncased}
\end{figure}

\begin{figure}[ht!]
    \centering
    \includegraphics[scale=0.3]{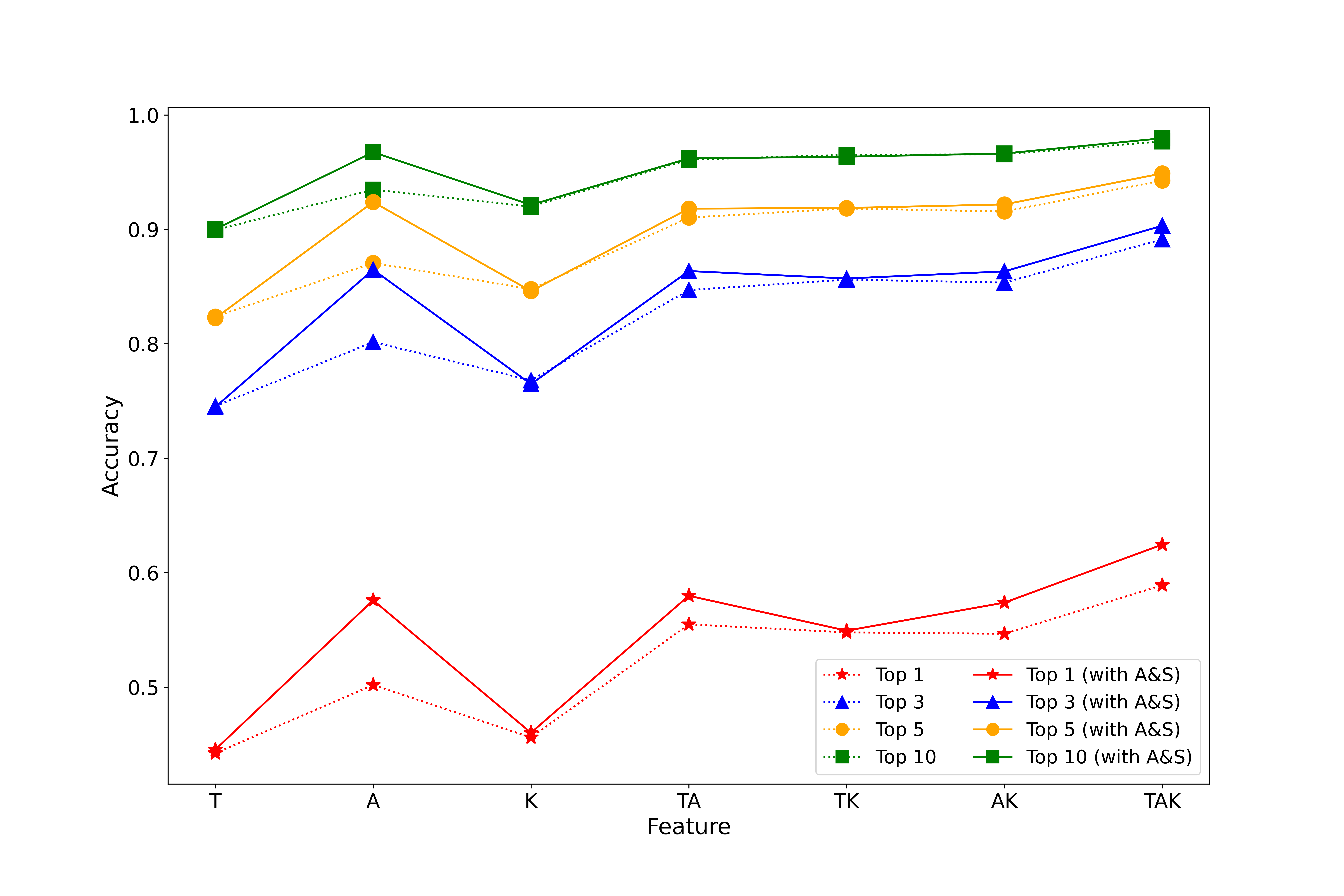}
    \caption{The performance of different features for the second approach that uses DistilBert cased. Here, we compare the performance by Accuracy@K (K = 1, 3, 5, 10) for two cases: without/with using “Aims and Scopes”}
    \label{fig: bert_cased}
\end{figure}

In summary, one can see comprehensively  the results in Figures \ref{fig: gru}, \ref{fig: bilstm},  \ref{fig: lstm},  \ref{fig: bert_uncased}, and \ref{fig: bert_cased}.

\begin{table}[]
    \centering
    \renewcommand{\arraystretch}{1.2}
    \caption{The performance of the baseline approach which is contributed by Son et al. in terms of Accuracy@K (K = 1, 3, 5, 10) for two cases: without/with using “Aims and Scopes”\cite{huynh2021fusion}.}
    \begin{tabular}{|c|*{4}{p{1.4cm}|}}
    \hline
    \textbf{Feature}  & \textbf{Top1} & \textbf{Top3} & \textbf{Top5} & \textbf{Top10}
    \\
    \hline
    T & 0.3542 & 0.6634 & 0.7561 & 0.8532
    \\
    \hline
    TS & \textbf{0.4015} & \textbf{0.6991} & \textbf{0.7971} & \textbf{0.8951}
    \\
    \hline
    K & 0.3933 & 0.7008 & 0.7919 & 0.8852
    \\
    \hline
    KS & \textbf{0.4284} & \textbf{0.7256} & \textbf{0.8189} & \textbf{0.9075}
    \\
    \hline
    A & 0.4691 & 0.7661 & 0.8482 & 0.9253
    \\
    \hline
    AS & \textbf{0.4770} & \textbf{0.7662} & \textbf{0.8488} & \textbf{0.9258}
    \\
    \hline
    TK & 0.4157 & 0.7315 & 0.8232 & 0.9084
    \\
    \hline
    TKS & \textbf{0.4475} & \textbf{0.7490} & \textbf{0.8302} & \textbf{0.9127}
    \\
    \hline
    TA & 0.4644 & 0.7613 & 0.8448 & 0.9233
    \\
    \hline
    TAS & \textbf{0.4828} & \textbf{0.7754} & \textbf{0.8536} & \textbf{0.9276}
    \\
    \hline
    AK & 0.4791 & 0.7730 & 0.8530 & 0.9273
    \\
    \hline
    AKS & \textbf{0.4951} & \textbf{0.7830} & \textbf{0.8602} & \textbf{0.9304}
    \\
    \hline
    TAK & 0.4852 & 0.7856 & 0.8624 & \textbf{0.9333}
    \\
    \hline
    TAKS & \textbf{0.5002} & \textbf{0.7889} & \textbf{0.8627} & 0.9323
    \\
    \hline
\end{tabular}
\centering
\label{table: old_result}
\end{table}



\begin{table}
\centering
\caption{The performance of our first approach that uses LSTM, BiLSTM, GRU, BiGRU as feature extractor. Here, we compare the performance by Accuracy@K (K = 1, 3, 5, 10) for two cases: without/with using “Aims and Scopes” (Part 1)}
\label{tab: result1}
\renewcommand{\arraystretch}{1.3}
\begin{tabular}{|l|l|l|l|l|l|} 
\hline
Method                   & Feature & Top1            & Top3            & Top5            & Top10            \\ 
\hline
\multirow{14}{*}{LSTM}   & TAKS    & 0.4825          & 0.7797          & 0.8641          & 0.9388           \\ 
\cline{2-6}
                         & TAK     & \textbf{0.4837} & \textbf{0.7851} & \textbf{0.8690} & \textbf{0.9420}  \\ 
\cline{2-6}
                         & TKS     & \textbf{0.4278} & \textbf{0.7301} & \textbf{0.8257} & \textbf{0.9178}  \\ 
\cline{2-6}
                         & TK      & 0.4056          & 0.7045          & 0.8053          & 0.9029           \\ 
\cline{2-6}
                         & AKS     & \textbf{0.4817} & 0.7771          & 0.8622          & 0.9373           \\ 
\cline{2-6}
                         & AK      & 0.4786          & \textbf{0.7780} & \textbf{0.8632} & \textbf{0.9392}  \\ 
\cline{2-6}
                         & TS      & \textbf{0.4001} & \textbf{0.6945} & \textbf{0.7980} & \textbf{0.9011}  \\ 
\cline{2-6}
                         & T       & 0.3471          & 0.6242          & 0.7276          & 0.8417           \\ 
\cline{2-6}
                         & KS      & \textbf{0.4210} & \textbf{0.7200} & \textbf{0.8181} & \textbf{0.9113}  \\ 
\cline{2-6}
                         & K       & 0.3757          & 0.6591          & 0.7632          & 0.8710           \\ 
\cline{2-6}
                         & TAS     & 0.4278          & 0.7301          & 0.8257          & 0.9178           \\ 
\cline{2-6}
                         & TA      & \textbf{0.4837} & \textbf{0.7851} & \textbf{0.8690} & \textbf{0.9420}  \\ 
\cline{2-6}
                         & AS      & \textbf{0.4654} & \textbf{0.7610} & 0.8494          & \textbf{0.9312}  \\ 
\cline{2-6}
                         & A       & 0.4615          & 0.7595          & \textbf{0.8497} & 0.9312           \\ 
\hline
\multirow{14}{*}{BiLSMT} & TAKS    & \textbf{0.4830} & \textbf{0.7807} & \textbf{0.8648} & 0.9398           \\ 
\cline{2-6}
                         & TAK     & 0.4782          & 0.7782          & 0.8645          & \textbf{0.9407}  \\ 
\cline{2-6}
                         & TKS     & \textbf{0.4296} & \textbf{0.7293} & \textbf{0.8261} & \textbf{0.9169}  \\ 
\cline{2-6}
                         & TK      & 0.4048          & 0.7030          & 0.8033          & 0.9013           \\ 
\cline{2-6}
                         & AKS     & \textbf{0.4817} & 0.7774          & \textbf{0.8624} & \textbf{0.9389}  \\ 
\cline{2-6}
                         & AK      & 0.4771          & \textbf{0.7784} & 0.8621          & 0.9388           \\ 
\cline{2-6}
                         & TS      & \textbf{0.4015} & \textbf{0.6982} & \textbf{0.8015} & \textbf{0.9021}  \\ 
\cline{2-6}
                         & T       & 0.3450          & 0.6220          & 0.7288          & 0.8394           \\ 
\cline{2-6}
                         & KS      & \textbf{0.4202} & \textbf{0.7219} & \textbf{0.8189} & \textbf{0.9112}  \\ 
\cline{2-6}
                         & K       & 0.3748          & 0.6622          & 0.7661          & 0.8724           \\ 
\cline{2-6}
                         & TAS     & 0.4296          & 0.7293          & 0.8261          & 0.9169           \\ 
\cline{2-6}
                         & TA      & \textbf{0.4685} & \textbf{0.7673} & \textbf{0.8548} & \textbf{0.9350}  \\ 
\cline{2-6}
                         & AS      & \textbf{0.4661} & \textbf{0.7570} & \textbf{0.8468} & 0.9290           \\ 
\cline{2-6}
                         & A       & 0.4606          & 0.7558          & 0.8460          & \textbf{0.9297}  \\
\hline
\end{tabular}
\end{table}

\begin{table}
\centering
\renewcommand{\arraystretch}{1.3}
\caption{The performance of our first approach using LSTM, BiLSTM, GRU, BiGRU as feature extractor. Here, we compare the performance by Accuracy@K (K = 1, 3, 5, 10) for two cases: without/with using “Aims and Scopes” (Part 2)}
\label{tab: result1_part2}
\begin{tabular}{|l|l|l|l|l|l|} 
\hline
Method                  & Feature & Top1            & Top3            & Top5            & Top10            \\ 
\hline
\multirow{14}{*}{GRU}   & TAKS    & \textbf{0.4898} & \textbf{0.7881} & \textbf{0.8707} & 0.9442           \\ 
\cline{2-6}
                        & TAK     & 0.4854          & 0.7864          & 0.8704          & \textbf{0.9444}  \\ 
\cline{2-6}
                        & TKS     & \textbf{0.4321} & \textbf{0.7347} & \textbf{0.8312} & \textbf{0.9199}  \\ 
\cline{2-6}
                        & TK      & 0.4087          & 0.7097          & 0.8106          & 0.9058           \\ 
\cline{2-6}
                        & AKS     & 0.4858          & 0.7847          & 0.8681          & 0.9417           \\ 
\cline{2-6}
                        & AK      & \textbf{0.4866} & \textbf{0.7899} & \textbf{0.8737} & \textbf{0.9450}  \\ 
\cline{2-6}
                        & TS      & \textbf{0.4059} & \textbf{0.7013} & \textbf{0.8048} & \textbf{0.9047}  \\ 
\cline{2-6}
                        & T       & 0.3473          & 0.6193          & 0.7258          & 0.8398           \\ 
\cline{2-6}
                        & KS      & \textbf{0.4259} & \textbf{0.7257} & \textbf{0.8223} & \textbf{0.9149}  \\ 
\cline{2-6}
                        & K       & 0.3757          & 0.6623          & 0.7649          & 0.8730           \\ 
\cline{2-6}
                        & TAS     & 0.4321          & 0.7347          & 0.8312          & 0.9199           \\ 
\cline{2-6}
                        & TA      & \textbf{0.4854} & \textbf{0.7864} & \textbf{0.8704} & \textbf{0.9444}  \\ 
\cline{2-6}
                        & AS      & \textbf{0.4747} & \textbf{0.7692} & \textbf{0.8565} & \textbf{0.9360}  \\ 
\cline{2-6}
                        & A       & 0.4689          & 0.7626          & 0.8537          & 0.9339           \\ 
\hline
\multirow{14}{*}{BiGRU} & TAKS    & \textbf{0.4895} & 0.7882          & 0.8716          & 0.9437           \\ 
\cline{2-6}
                        & TAK     & 0.4881          & \textbf{0.7907} & \textbf{0.8737} & \textbf{0.9451}  \\ 
\cline{2-6}
                        & TKS     & \textbf{0.4321} & \textbf{0.7346} & \textbf{0.8308} & \textbf{0.9218}  \\ 
\cline{2-6}
                        & TK      & 0.4071          & 0.7095          & 0.8093          & 0.9056           \\ 
\cline{2-6}
                        & AKS     & \textbf{0.4898} & \textbf{0.7867} & 0.8704          & 0.9436           \\ 
\cline{2-6}
                        & AK      & 0.4877          & 0.7869          & \textbf{0.8711} & \textbf{0.9449}  \\ 
\cline{2-6}
                        & TS      & \textbf{0.4009} & \textbf{0.6957} & \textbf{0.7979} & \textbf{0.9005}  \\ 
\cline{2-6}
                        & T       & 0.3485          & 0.6247          & 0.7309          & 0.8427           \\ 
\cline{2-6}
                        & KS      & \textbf{0.4204} & \textbf{0.7230} & \textbf{0.8216} & \textbf{0.9138}  \\ 
\cline{2-6}
                        & K       & 0.3806          & 0.6644          & 0.7665          & 0.8726           \\ 
\cline{2-6}
                        & TAS     & \textbf{0.4321} & \textbf{0.7346} & \textbf{0.8308} & \textbf{0.9218}  \\ 
\cline{2-6}
                        & TA      & 0.4881          & 0.7907          & 0.8737          & 0.9451           \\ 
\cline{2-6}
                        & AS      & \textbf{0.4718} & \textbf{0.7678} & \textbf{0.8550} & \textbf{0.9345}  \\ 
\cline{2-6}
                        & A       & 0.4692          & 0.7628          & 0.8526          & 0.9333           \\
\hline
\end{tabular}
\end{table}

\begin{table}
\centering
\renewcommand{\arraystretch}{1.3}
\caption{The performance of our second approach using DistilBert cased and uncased. Here, we compare the performance by Accuracy@K (K = 1, 3, 5, 10) for two cases: without/with using “Aims and Scopes”}
\label{tab: result2}
\begin{tabular}{|l|l|l|l|l|l|} 
\hline
Method                                     & Feature & Top1            & Top3            & Top5            & Top10            \\ 
\hline
\multirow{14}{*}{DistilBert+CNN1D+uncased} & TAKS    & \textbf{0.5537} & \textbf{0.8409} & \textbf{0.901}  & \textbf{0.9524}  \\ 
\cline{2-6}
                                           & TAK     & 0.5503          & 0.8398          & 0.8959          & 0.9479           \\ 
\cline{2-6}
                                           & TKS     & \textbf{0.4564} & \textbf{0.7645} & 0.8427          & \textbf{0.9179}  \\ 
\cline{2-6}
                                           & TK      & 0.4537          & 0.7638          & \textbf{0.8432} & 0.9174           \\ 
\cline{2-6}
                                           & AKS     & \textbf{0.5437} & \textbf{0.8396} & 0.8948          & 0.9456           \\ 
\cline{2-6}
                                           & AK      & 0.5418          & 0.8438          & \textbf{0.9117} & \textbf{0.9620}  \\ 
\cline{2-6}
                                           & TS      & 0.3851          & 0.6806          & 0.7702          & 0.8627           \\ 
\cline{2-6}
                                           & T       & \textbf{0.3867} & \textbf{0.6843} & \textbf{0.7742} & \textbf{0.8660}  \\ 
\cline{2-6}
                                           & KS      & \textbf{0.4237} & 0.7271          & 0.8070          & 0.8876           \\ 
\cline{2-6}
                                           & K       & 0.4232          & \textbf{0.7282} & \textbf{0.8091} & \textbf{0.8911}  \\ 
\cline{2-6}
                                           & TAS     & \textbf{0.5294} & \textbf{0.8219} & 0.8850          & 0.9417           \\ 
\cline{2-6}
                                           & TA      & 0.5253          & 0.8203          & \textbf{0.8879} & \textbf{0.9447}  \\ 
\cline{2-6}
                                           & AS      & \textbf{0.5279} & \textbf{0.8186} & \textbf{0.8856} & \textbf{0.9440}  \\ 
\cline{2-6}
                                           & A       & 0.4458          & 0.7538          & 0.8361          & 0.9136           \\ 
\hline
\multirow{14}{*}{DistilBert+CNN1D+cased}   & TAKS    & \textbf{0.6246} & \textbf{0.9032} & \textbf{0.9489} & \textbf{0.9796}  \\ 
\cline{2-6}
                                           & TAK     & \textbf{0.5891} & \textbf{0.8913} & \textbf{0.9427} & \textbf{0.9769}  \\ 
\cline{2-6}
                                           & TKS     & \textbf{0.5494} & \textbf{0.8571} & \textbf{0.9187} & 0.9635           \\ 
\cline{2-6}
                                           & TK      & 0.5479          & 0.8561          & 0.9183          & \textbf{0.9651}  \\ 
\cline{2-6}
                                           & AKS     & \textbf{0.5739} & \textbf{0.8633} & \textbf{0.9218} & \textbf{0.9663}  \\ 
\cline{2-6}
                                           & AK      & 0.5467          & 0.8535          & 0.9156          & 0.9656           \\ 
\cline{2-6}
                                           & TS      & \textbf{0.4455} & \textbf{0.7448} & 0.8225          & \textbf{0.9001}  \\ 
\cline{2-6}
                                           & T       & 0.4423          & 0.7456          & \textbf{0.8238} & 0.8994           \\ 
\cline{2-6}
                                           & KS      & \textbf{0.4602} & 0.7648          & 0.8461          & \textbf{0.9214}  \\ 
\cline{2-6}
                                           & K       & 0.4561          & \textbf{0.7681} & \textbf{0.8478} & 0.9199           \\ 
\cline{2-6}
                                           & TAS     & \textbf{0.5799} & \textbf{0.8636} & \textbf{0.9181} & \textbf{0.9620}  \\ 
\cline{2-6}
                                           & TA      & 0.5549          & 0.8470          & 0.9103          & \textbf{0.9609}  \\ 
\cline{2-6}
                                           & AS      & \textbf{0.576}  & \textbf{0.8648} & \textbf{0.9239} & 0.9674           \\ 
\cline{2-6}
                                           & A       & 0.5020          & 0.8016          & 0.8707          & 0.9347           \\
\hline
\end{tabular}
\end{table}

\begin{figure}[ht!]
    \centering
    \includegraphics[scale=0.3]{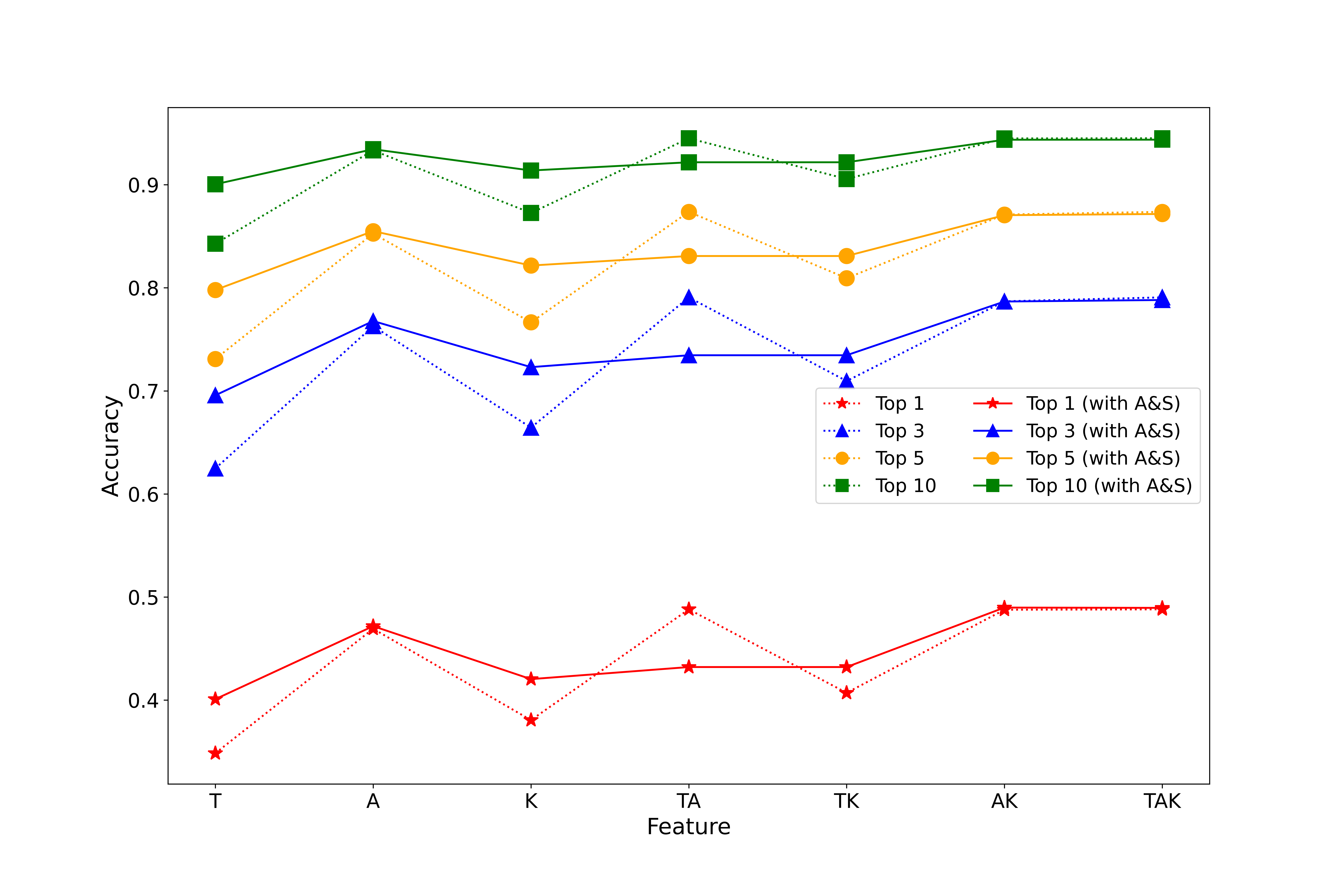}
    \caption{The performance of different features for the first approach using BiGRU as a feature extractor. Here, we compare the performance by Accuracy@K (K = 1, 3, 5, 10) for two cases: without/with using “Aims and Scopes”}
    \label{fig: bigru}
\end{figure}

\section{Conclusion and Future Works}
From a technique used, FastText as an embedding layer and Convolution 1D as a feature extractor for the problem of paper submission recommendation system in previous our paper \cite{huynh2021fusion} at IEA/AIE 2021. In this paper, we have contributed two different approaches; one extends from previous our article by using variations of Recurrent Neural Network instead of Convolution 1D as a feature extractor. The other utilizes DistilBert for embedding layer instead of FastText for Title, Abstract, Keyword, and Aim \& Scope in our old paper \cite{huynh2021fusion}. Furthermore, we utilize multi Convolution 1D with different kernel sizes to capture various semantics of the input. The result shows that the performance of this technique outperforms all previous methods.
Interestingly, we also introduce a new method for calculating the similarity of Aim \& Scope. Combining these methods has shown the best result compared to other techniques. 

In the future, we plan to experiment on a larger dataset and the various publisher instead of Springer and use other ways to calculate similarity scores. Furthermore, to reinforce the performance of the model performance not only accuracy but speed, we extend our experiment on using Transformer with the Fourier Transform as an alternative to Self-Attention according to the new result from the Google Research team \cite{leethorp2021fnet}. And by applying the Mixture of Experts (MoE) inside the model architecture inspired by Nan Du et al. \cite{du2021glam}, we expect that will enhance the completeness as well as the ability of the model's architecture (increase the total parameters but reduce the running parameters in an inference process).

\section*{Acknowledgement}
Son Huynh Thanh was funded by Vingroup JSC and supported by the Master, PhD Scholarship Programme of Vingroup Innovation Foundation (VINIF), Institute of Big Data, code VINIF.2021.ThS.18

\nocite{*}
\bibliographystyle{spbasic}      
\bibliography{reference}   

\begin{thebibliography}{31}
\providecommand{\natexlab}[1]{#1}
\providecommand{\url}[1]{{#1}}
\providecommand{\urlprefix}{URL }
\expandafter\ifx\csname urlstyle\endcsname\relax
  \providecommand{\doi}[1]{DOI~\discretionary{}{}{}#1}\else
  \providecommand{\doi}{DOI~\discretionary{}{}{}\begingroup
  \urlstyle{rm}\Url}\fi
\providecommand{\eprint}[2][]{\url{#2}}

\bibitem[{Bojanowski et~al.(2017)Bojanowski, Grave, Joulin, and
  Mikolov}]{fasttext}
Bojanowski P, Grave E, Joulin A, Mikolov T (2017) Enriching word vectors with
  subword information. Transactions of the Association for Computational
  Linguistics 5:135--146

\bibitem[{Bromley et~al.(1993)Bromley, Guyon, LeCun, S\"{a}ckinger, and
  Shah}]{siamese2}
Bromley J, Guyon I, LeCun Y, S\"{a}ckinger E, Shah R (1993) Signature
  verification using a "siamese" time delay neural network. In: Proceedings of
  the 6th International Conference on Neural Information Processing Systems,
  Morgan Kaufmann Publishers Inc., San Francisco, CA, USA, NIPS'93, p 737–744

\bibitem[{Buciluundefined et~al.(2006)Buciluundefined, Caruana, and
  Niculescu-Mizil}]{distilation1}
Buciluundefined C, Caruana R, Niculescu-Mizil A (2006) Model compression. In:
  Proceedings of the 12th ACM SIGKDD International Conference on Knowledge
  Discovery and Data Mining, Association for Computing Machinery, New York, NY,
  USA, KDD '06, p 535–541, \doi{10.1145/1150402.1150464},
  \urlprefix\url{https://doi.org/10.1145/1150402.1150464}

\bibitem[{Chicco(2021)}]{siamese1}
Chicco D (2021) Siamese Neural Networks: An Overview, Springer US, New York,
  NY, pp 73--94. \doi{10.1007/978-1-0716-0826-5_3},
  \urlprefix\url{https://doi.org/10.1007/978-1-0716-0826-5_3}

\bibitem[{Cho et~al.(2014)Cho, van Merri{\"e}nboer, Gulcehre, Bahdanau,
  Bougares, Schwenk, and Bengio}]{gru}
Cho K, van Merri{\"e}nboer B, Gulcehre C, Bahdanau D, Bougares F, Schwenk H,
  Bengio Y (2014) Learning phrase representations using {RNN}
  encoder{--}decoder for statistical machine translation. In: Proceedings of
  the 2014 Conference on Empirical Methods in Natural Language Processing
  ({EMNLP}), Association for Computational Linguistics, Doha, Qatar, pp
  1724--1734, \doi{10.3115/v1/D14-1179},
  \urlprefix\url{https://aclanthology.org/D14-1179}

\bibitem[{Devlin et~al.(2018)Devlin, Chang, Lee, and Toutanova}]{bert}
Devlin J, Chang MW, Lee K, Toutanova K (2018) Bert: Pre-training of deep
  bidirectional transformers for language understanding.
  \urlprefix\url{http://arxiv.org/abs/1810.04805}, cite
  arxiv:1810.04805Comment: 13 pages

\bibitem[{Du et~al.(2021)Du, Huang, Dai, Tong, Lepikhin, Xu, Krikun, Zhou, Yu,
  Firat, Zoph, Fedus, Bosma, Zhou, Wang, Wang, Webster, Pellat, Robinson,
  Meier-Hellstern, Duke, Dixon, Zhang, Le, Wu, Chen, and Cui}]{du2021glam}
Du N, Huang Y, Dai AM, Tong S, Lepikhin D, Xu Y, Krikun M, Zhou Y, Yu AW, Firat
  O, Zoph B, Fedus L, Bosma M, Zhou Z, Wang T, Wang YE, Webster K, Pellat M,
  Robinson K, Meier-Hellstern K, Duke T, Dixon L, Zhang K, Le QV, Wu Y, Chen Z,
  Cui C (2021) Glam: Efficient scaling of language models with
  mixture-of-experts. \eprint{2112.06905}

\bibitem[{Han et~al.(2012)Han, Kamber, and Pei}]{HAN201239}
Han J, Kamber M, Pei J (2012) Section 2.4.7 cosine simularity in chapter 2 -
  getting to know your data. In: Han J, Kamber M, Pei J (eds) Data Mining
  (Third Edition), third edition edn, The Morgan Kaufmann Series in Data
  Management Systems, Morgan Kaufmann, Boston, pp 77--78,
  \doi{https://doi.org/10.1016/B978-0-12-381479-1.00002-2},
  \urlprefix\url{https://www.sciencedirect.com/science/article/pii/B9780123814791000022}

\bibitem[{He et~al.(2016)He, Zhang, Ren, and Sun}]{resnet}
He K, Zhang X, Ren S, Sun J (2016) Deep residual learning for image
  recognition. 2016 IEEE Conference on Computer Vision and Pattern Recognition
  (CVPR) pp 770--778

\bibitem[{Hinton et~al.(2015)Hinton, Vinyals, and Dean}]{distilation2}
Hinton G, Vinyals O, Dean J (2015) Distilling the knowledge in a neural
  network. \eprint{1503.02531}

\bibitem[{Hochreiter and Schmidhuber(1997)}]{lstm}
Hochreiter S, Schmidhuber J (1997) Long short-term memory. Neural computation
  9:1735--80, \doi{10.1162/neco.1997.9.8.1735}

\bibitem[{Huynh et~al.(2020)Huynh, Huynh, Nguyen, Cuong, and
  Nguyen}]{ieaaie2020}
Huynh ST, Huynh PT, Nguyen DH, Cuong DV, Nguyen BT (2020) S2rscs: An efficient
  scientific submission recommendation system for computer science. In: Fujita
  H, Fournier-Viger P, Ali M, Sasaki J (eds) Trends in Artificial Intelligence
  Theory and Applications. Artificial Intelligence Practices, Springer
  International Publishing, Cham, pp 186--198

\bibitem[{Huynh et~al.(2021)Huynh, Dang, Huynh, Nguyen, and
  Nguyen}]{huynh2021fusion}
Huynh ST, Dang N, Huynh PT, Nguyen DH, Nguyen BT (2021) A fusion approach for
  paper submission recommendation system. In: International Conference on
  Industrial, Engineering and Other Applications of Applied Intelligent
  Systems, Springer, pp 72--83

\bibitem[{Iandola et~al.(2016)Iandola, Han, Moskewicz, Ashraf, Dally, and
  Keutzer}]{iandola2016squeezenet}
Iandola FN, Han S, Moskewicz MW, Ashraf K, Dally WJ, Keutzer K (2016)
  Squeezenet: Alexnet-level accuracy with 50x fewer parameters and <0.5mb model
  size. \eprint{1602.07360}

\bibitem[{Jordan(1997)}]{rnn2}
Jordan MI (1997) Chapter 25 - serial order: A parallel distributed processing
  approach. In: Donahoe JW, {Packard Dorsel} V (eds) Neural-Network Models of
  Cognition, Advances in Psychology, vol 121, North-Holland, pp 471--495,
  \doi{https://doi.org/10.1016/S0166-4115(97)80111-2},
  \urlprefix\url{https://www.sciencedirect.com/science/article/pii/S0166411597801112}

\bibitem[{Kim(2014)}]{kim2014convolutional}
Kim Y (2014) Convolutional neural networks for sentence classification.
  \eprint{1408.5882}

\bibitem[{Kingma and Ba(2015)}]{adam}
Kingma DP, Ba J (2015) Adam: {A} method for stochastic optimization. In: Bengio
  Y, LeCun Y (eds) 3rd International Conference on Learning Representations,
  {ICLR} 2015, San Diego, CA, USA, May 7-9, 2015, Conference Track Proceedings,
  \urlprefix\url{http://arxiv.org/abs/1412.6980}

\bibitem[{Klamma et~al.(2009)Klamma, Pham, and Cao}]{original1}
Klamma R, Pham MC, Cao Y (2009) You never walk alone: Recommending academic
  events based on social network analysis. In: Zhou J (ed) Complex Sciences,
  First International Conference, Complex 2009, Shanghai, China, February
  23-25, 2009. Revised Papers, Part 1, Springer, Lecture Notes of the Institute
  for Computer Sciences, Social Informatics and Telecommunications Engineering,
  vol~4, pp 657--670, \doi{10.1007/978-3-642-02466-5\_64},
  \urlprefix\url{https://doi.org/10.1007/978-3-642-02466-5\_64}

\bibitem[{Krizhevsky et~al.(2012)Krizhevsky, Sutskever, and Hinton}]{alexnet}
Krizhevsky A, Sutskever I, Hinton GE (2012) Imagenet classification with deep
  convolutional neural networks. In: Proceedings of the 25th International
  Conference on Neural Information Processing Systems - Volume 1, Curran
  Associates Inc., Red Hook, NY, USA, NIPS'12, p 1097–1105

\bibitem[{Lecun et~al.(1998)Lecun, Bottou, Bengio, and Haffner}]{726791}
Lecun Y, Bottou L, Bengio Y, Haffner P (1998) Gradient-based learning applied
  to document recognition. Proceedings of the IEEE 86(11):2278--2324,
  \doi{10.1109/5.726791}

\bibitem[{Lee-Thorp et~al.(2021)Lee-Thorp, Ainslie, Eckstein, and
  Ontanon}]{leethorp2021fnet}
Lee-Thorp J, Ainslie J, Eckstein I, Ontanon S (2021) Fnet: Mixing tokens with
  fourier transforms. \eprint{2105.03824}

\bibitem[{Liu and Deng(2015)}]{vgg}
Liu S, Deng W (2015) Very deep convolutional neural network based image
  classification using small training sample size. In: 2015 3rd IAPR Asian
  Conference on Pattern Recognition (ACPR), pp 730--734,
  \doi{10.1109/ACPR.2015.7486599}

\bibitem[{Luong et~al.(2012{\natexlab{a}})Luong, Huynh, Gauch, Do, and
  Hoang}]{luong1}
Luong H, Huynh T, Gauch S, Do L, Hoang K (2012{\natexlab{a}}) Publication venue
  recommendation using author network's publication history. In: Pan JS, Chen
  SM, Nguyen NT (eds) Intelligent Information and Database Systems, Springer
  Berlin Heidelberg, Berlin, Heidelberg, pp 426--435

\bibitem[{Luong et~al.(2012{\natexlab{b}})Luong, Huynh, Gauch, Do, and
  Hoang}]{luong}
Luong H, Huynh T, Gauch S, Do L, Hoang K (2012{\natexlab{b}}) Publication venue
  recommendation using author network's publication history. In: Pan JS, Chen
  SM, Nguyen NT (eds) Intelligent Information and Database Systems, Springer
  Berlin Heidelberg, Berlin, Heidelberg, pp 426--435

\bibitem[{{Medvet} et~al.(2014){Medvet}, {Bartoli}, and {Piccinin}}]{Medvet}
{Medvet} E, {Bartoli} A, {Piccinin} G (2014) Publication venue recommendation
  based on paper abstract. In: 2014 IEEE 26th International Conference on Tools
  with Artificial Intelligence, pp 1004--1010, \doi{10.1109/ICTAI.2014.152}

\bibitem[{Nguyen et~al.(2021)Nguyen, Huynh, Huynh, Dinh, and Nguyen}]{Sofsem}
Nguyen D, Huynh S, Huynh P, Dinh CV, Nguyen BT (2021) S2cft: A new approach for
  paper submission recommendation. In: SOFSEM 2021: Theory and Practice of
  Computer Science, Springer International Publishing, Cham, pp 563--573

\bibitem[{Rumelhart et~al.(1986)Rumelhart, Hinton, and Williams}]{rnn1}
Rumelhart DE, Hinton GE, Williams RJ (1986) Learning Internal Representations
  by Error Propagation, MIT Press, Cambridge, MA, USA, p 318–362

\bibitem[{Sanh et~al.(2019)Sanh, Debut, Chaumond, and Wolf}]{ditlbert}
Sanh V, Debut L, Chaumond J, Wolf T (2019) Distilbert, a distilled version of
  bert: smaller, faster, cheaper and lighter. ArXiv abs/1910.01108

\bibitem[{Szegedy et~al.(2016)Szegedy, Ioffe, and
  Vanhoucke}]{DBLP:journals/corr/SzegedyIV16}
Szegedy C, Ioffe S, Vanhoucke V (2016) Inception-v4, inception-resnet and the
  impact of residual connections on learning. CoRR abs/1602.07261,
  \urlprefix\url{http://arxiv.org/abs/1602.07261}, \eprint{1602.07261}

\bibitem[{Wang et~al.(2018)Wang, Liang, Xu, Feng, and Guan}]{WANG20181}
Wang D, Liang Y, Xu D, Feng X, Guan R (2018) A content-based recommender system
  for computer science publications. Knowledge-Based Systems 157:1 -- 9,
  \doi{https://doi.org/10.1016/j.knosys.2018.05.001},
  \urlprefix\url{http://www.sciencedirect.com/science/article/pii/S0950705118302107}

\bibitem[{Zhang et~al.(2016)Zhang, Zhao, and LeCun}]{zhang2016characterlevel}
Zhang X, Zhao J, LeCun Y (2016) Character-level convolutional networks for text
  classification. \eprint{1509.01626}

\end{thebibliography}

\end{document}